\documentclass[aps,pra,graphicx,9pt,twocolumn,showkeys,superscriptaddress]{revtex4-2}
\usepackage[colorlinks,linkcolor=blue,urlcolor=blue,anchorcolor=blue,citecolor=blue]{hyperref}
\usepackage{amsmath}
\usepackage{graphicx}
\usepackage{dcolumn}
\usepackage{mathrsfs}
\usepackage{amssymb}
\usepackage{amsfonts,multirow}
\usepackage{bm,changes}
\usepackage{setspace} 
\usepackage{color}
\usepackage{upgreek}
\usepackage[thicklines]{cancel}
\newcommand{\Rmnum}[1]{\expandafter\@slowromancap\romannumeral #1@}
\begin{document}
\title{Quantum-enhanced estimation of signal field amplitudes with critical squeezed states of photonic modes}

\author{Ken Chen}
\author{Jia-Hao L\"{u}}
\author{Wen Ning} \email{ningw@fzu.edu.cn}
\address{Fujian Key Laboratory of Quantum Information and Quantum\\
Optics, College of Physics and Information Engineering,\\
Fuzhou University, Fuzhou, Fujian 350108, China}

\author{Zhen-Biao Yang} \email{zbyang@fzu.edu.cn}
\author{Shi-Biao Zheng} \email{t96034@fzu.edu.cn}
\address{Fujian Key Laboratory of Quantum Information and Quantum\\
Optics, College of Physics and Information Engineering,\\
Fuzhou University, Fuzhou, Fujian 350108, China}
\address{Hefei National Laboratory, Hefei 230088, China}

\begin{abstract}
Critical phenomena of quantum systems offer a promising strategy to improve
measurement precision. So far, many criticality-enhanced quantum
metrological schemes have been proposed by using the adiabatically evolved
photonic states of composite systems involving a qubit and a field
interacting with each other. These schemes focus on the measurement of the
system's inherent frequencies. We here propose a criticality-enhanced
quantum sensing protocol, aiming to estimate the amplitude of an external
signal field with the interacting qubit-photon system. The signal field is
coupled to the photonic mode, so that the composite system has a unique dark
state, where the photonic mode follows a squeezed vacuum state. The
information about the signal field amplitude is encoded in photon number or one quadrature of
the quantized photonic mode, which exhibits a divergent behavior near the
critical point. The measurement precision can approach the Heisenberg limit
with respect to the time to encode the signal and the photon number of the
field mode.
\end{abstract}


\maketitle

\section{INTRODUCTION}
\bigskip One of the most fundamental principles of quantum mechanics is the
Heisenberg uncertainty relation \cite{Robertson1929Uncertainty}, which states that two conjugate
observables of a quantum system cannot have definite values at the same
time. Such an uncertainty originates from intrinsic wave-particle duality,
and hence cannot be overcome by improving measurement techniques. 
This is exemplified by quasiclassical coherent fields, which occupy a minimum area of $1/2$ in the phase space defined by two dimensionless quadratures. Their phase-space extent is fully characterized by the quantum fluctuations of these quadratures. This inherent uncertainty imposes a lower
bound on the smallest translation in phase space detectable by classical
means \cite{Toscano2006Sub}. This bound, referred to as the standard quantum limit (SQL) or
the shot-noise limit, is independent of the photon number of the field.

To surpass SQL, nonclassical states are required. A paradigmatic example is the squeezed state, which reduces quantum fluctuations in one quadrature below the vacuum level at the expense of increased fluctuations in the other one \cite{Braun2018Quantum}. When the squeezing
is sufficiently strong, the precision for measuring a small displacement in
phase space approximately scales inversely with the square root of the
photon number, approaching the Heisenberg limit (HL). The HL for estimating
the phase-space displacement can also be approached with cat states \cite{Toscano2006Sub,Penasa2016Measurement}
and Fock states \cite{Deng2024Quantum}. The displacement can be induced by coupling a classical
signal field to the quantized photonic field, which acts as the sensor. This
implies that the amplitude of the signal field can be estimated by probing
the resulting phase-space displacement of the sensing field. Such
nonclassical states can also serve as a resource for enhancing measurement
of the frequency of the bosonic mode itself \cite{Vlastakis2013Deterministically, Pan2025Realization, Zheng2025Quantum, Wang2019Heisenberg, McCormick2019Quantum}. 

Quantum critical phenomena represent an alternative promising resource for
realizing quantum-enhanced metrology. In proximity of a quantum phase
transition, a quantum system exhibits an ultra-high sensitivity to a tiny
change of the control parameter of the Hamiltonian. The past two decades
have witnessed significant theoretical advances on critical quantum
metrology \cite{You2007Fidelity,Zanardi2008Quantum,Invernizzi2008Optimal,Albuquerque2010Quantum,Gammelmark2011Phase,Tsang2013Quantum,Ivanov2013Adiabatic,Salvatori2014Quantum,Bina2016Dicke,FernandezLorenzo2017Quantum,Raghunandan2018High,Frerot2018Quantum,Heugel2019Quantum,Wald2020out,Garbe2022Critical,Mihailescu2024Multiparameter,Ilias2022Criticality,DiCandia2023Critical,Chen2024Critical}. 
So far, criticality-enhanced quantum sensing has been
demonstrated in several quantum systems, including nuclear magnetic
resonance \cite{Liu2021Experimental} and Rydberg atomic gases \cite{Ding2022Enhanced}. Unlike the conventional
scenarios, critical metrological methods can reach an apparent
super-Heisenberg scaling in terms of the number of elements involved in the
critical system \cite{Rams2018Limits}. However, the HL is recovered when the time needed for
preparation of the quantum state that encodes the parameter is taken as
additional ingredient or resource. During the past few years, several
quantum sensing protocols have been proposed by exploiting the critical
behaviors of light-matter systems \cite{Garbe2020Critical,Chu2021Dynamic,Zhu2023Criticality,Lue2022Critical,Lue2023Quantum,Gietka2022Understanding}. In particular, it has been shown
that the Rabi model \cite{Garbe2020Critical,Chu2021Dynamic,Zhu2023Criticality} and the parametrically-driven Jaynes-Cummings model (JCM) \cite{Lue2022Critical} near the superradiant phase transition can serve as a critical
quantum sensor for estimating the system frequencies, with the measurement
precision approaching the Heisenberg scaling with respect to both the time
and photon number.

We have experimentally demonstrated a criticality-enhanced metrological protocol based on the driven JCM. In such a model, a photonic mode resonantly interacts with a qubit and is simultaneously coupled to a signal field whose amplitude is to be estimated \cite{Lue2026Critical}. 
In the scheme, the qubit's
excitation number serves as the indicator for extracting the information
about the amplitude of the signal field, which is robust against dissipation
and non-adiabaticity. The Fisher information exhibits a Heisenberg scaling
with respect to the evolution time, but does not show a direct dependence on
the photon number of the field mode. In the present work, we propose an
alternative critical quantum metrological scheme based on such a model.
Unlike the previous scheme, the signal is encoded in photon number or one quadrature of the
photonic mode, and the corresponding Fisher information presents a
Heisenberg scaling with respect to both the time and photon number as long
as the decoherence effects can be neglected.

\section{Critical quantum sensing with THE DRIVEN JCM}

The critical quantum system is realized by the driven JCM, 
for which a qubit is coupled to a quantized photonic mode that is pumped by an external drive. We consider the case where the qubit, the photonic mode and the drive are on resonance. 
In the interaction picture, the dynamics of the system is governed by the interaction Hamiltonian ($\hbar = 1$ is set)
\begin{equation}
H_{I} =\Omega \left[ a^{\dagger }\left\vert g\right\rangle \left\langle
e\right\vert +a\left\vert e\right\rangle \left\langle g\right\vert  +\eta (a^{\dagger }+a)/2\right], \label{eq:H} 
\end{equation}
where $\left\vert g\right\rangle$ and $\left\vert e\right\rangle$ label the ground and excited states of the qubit, $a^{\dagger}$ and $a$ denote the creation and annihilation operators of the photonic mode, $\Omega$ characterizes the qubit-photon coupling strength, and $\eta$ corresponds to the rescaled amplitude of the drive.
In the undriven limit ($\eta = 0$), the Hamiltonian reduces to the standard JCM, which exhibits a continuous U(1) symmetry stemming from the conservation of the total excitation number,
$N=a^{\dagger }a+\left\vert e\right\rangle \left\langle e\right\vert$.
This conserved quantity partitions the full Hilbert space into an infinite collection of invariant subspaces, each labeled by a fixed integer $n \geq 0$. For $n \geq 1$, the eigenstates within each subspace are superpositions of the form $(\left\vert n-1\right\rangle \left\vert e\right\rangle \pm \left\vert n\right\rangle \left\vert g\right\rangle )/\sqrt{2}$, with corresponding eigenenergies $\pm \sqrt{n}\,\Omega$. Specifically, the $n=0$ sector supports only one state, $\left\vert 0\right\rangle \left\vert g\right\rangle$, which has zero energy and is decoupled from the dynamics; this is the well-known dark state of the standard JCM.

When the photonic mode is linearly driven, the U(1) symmetry is explicitly broken by the linear coupling between the signal field and the photonic mode. 
Nevertheless, for driving strengths satisfying $\eta<1$, the characteristic $\sqrt{n}$ scaling of the JCM ladder is preserved, and specifically, the eigenvalues become \begin{equation}
    E_{n,\pm}=\pm \sqrt{n}\Omega (1-\eta ^{2})^{3/4}.
\end{equation}
The associated eigenstates are no longer simple Fock-state superpositions but are instead of the form that is generated by applying squeezing and displacement transformations to the bare JCM basis:
\begin{equation}
\left\vert \psi _{n,\pm }\right\rangle =\frac{1}{\sqrt{2}}S(r )D(\alpha
_{n,\pm })\left(\left\vert n-1\right\rangle \left\vert \Phi _{1}\right\rangle \pm
\left\vert n\right\rangle \left\vert \Phi _{0}\right\rangle \right),
\end{equation}
where 
\begin{equation}
    \begin{aligned}
\left\vert \Phi _{0}\right\rangle ={}&C\left\vert g\right\rangle -\sqrt{
1-C^{2}}\left\vert e\right\rangle , \\
\left\vert \Phi _{1}\right\rangle ={}&C\left\vert e\right\rangle -
\sqrt{1-C^{2}}\left\vert g\right\rangle,  
    \end{aligned}
\end{equation}
with $C=(1+\sqrt{1-\eta ^{2}})^{1/2}/\sqrt{2}$. 
Here, $S(r )=\exp [r(a^{2}- a^{\dagger 2})/2]$ and $D(\alpha _{n,\pm })=\exp (\alpha _{n,\pm }a^{\dagger }-\alpha _{n,\pm }^{\ast }a)$ denote the squeezing and displacement operators acting on the photonic field, respectively, with the parameters given by $r =\frac{1}{4}\ln (1-\eta ^{2})$ and $\alpha _{n,\pm }=\mp \sqrt{n}\eta$.

Notably, the dark state, which is reshaped by the drive, now takes the form
\begin{equation}
    \left\vert \psi _{0}\right\rangle =\left\vert \phi_r \right\rangle \left\vert
    \Phi _{0}\right\rangle ,
\end{equation}
where $\left\vert \phi_r \right\rangle = S(r)\left\vert 0\right\rangle$ denotes a squeezed vacuum state of the photonic mode. Thus, the new dark state is a separable state comprising a qubit-state superposition $\left\vert \Phi_0 \right\rangle$ and a nonclassical field state $\left\vert \phi_r \right\rangle$, both of which depend continuously on the drive amplitude $\eta$. This $\eta$-dependent structure enables the critical enhancement exploited in our sensing protocol.

\begin{figure}[t]
    \centering
    \includegraphics[width=\linewidth]{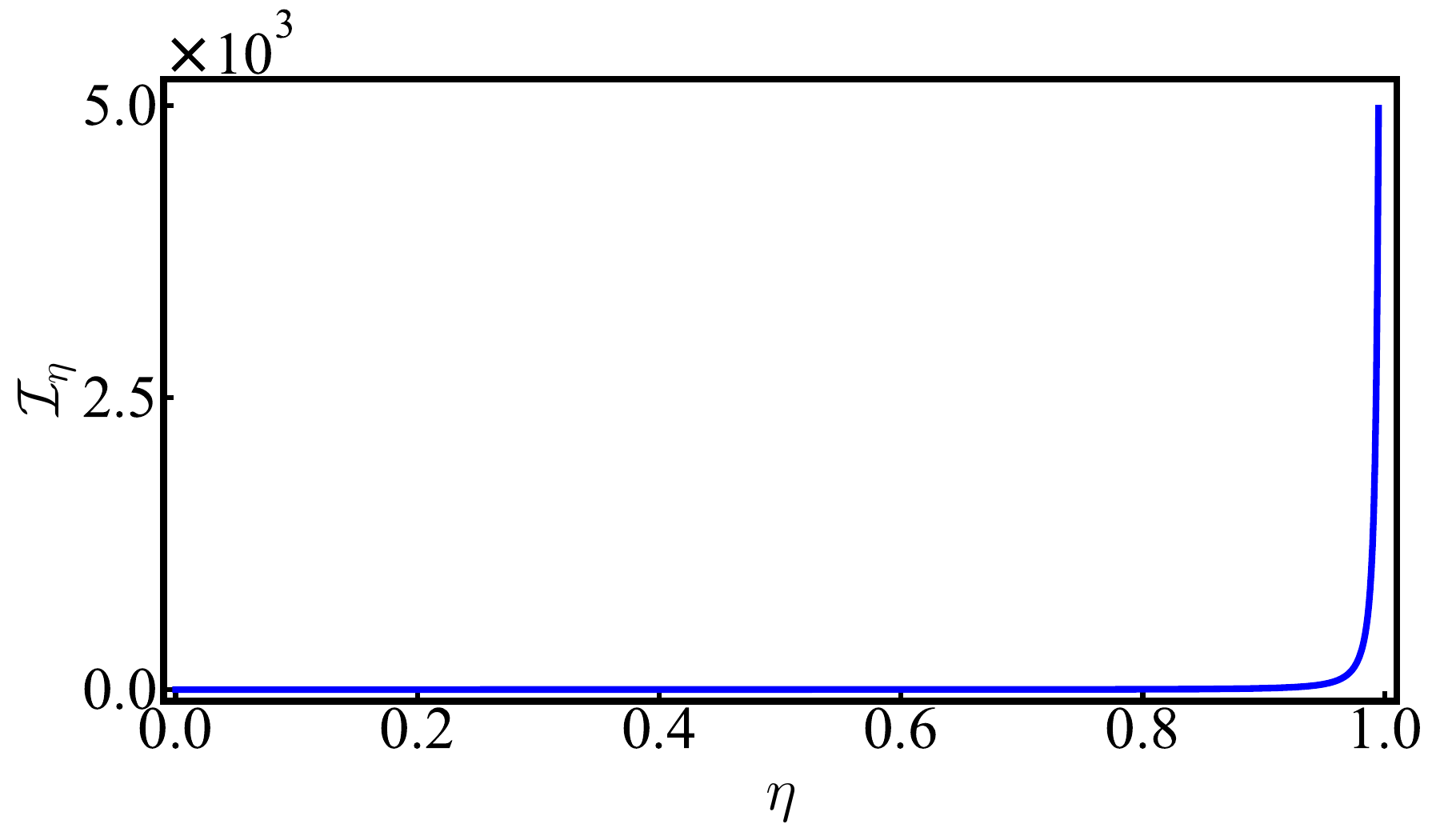}
    \caption{
    The QFI $\mathcal{I}_\eta$ versus the control parameter $\eta$. 
    The horizontal axis is truncated at the point where $\eta = 0.995$. 
    As $\eta$ approaches $1$, $\mathcal{I}_\eta$ exhibits a pronounced divergence.    
    }
    \label{fig:inverted_variance_eta}
\end{figure}

Our goal is to achieve high-precision estimation of the rescaled amplitude $\eta$ of the signal field by performing measurements on the ground state of the photonic mode.  
Because this ground state is pure, the quantum Fisher information (QFI) can be evaluated analytically using the standard expression for pure states:
\begin{equation}
    \begin{aligned}   
    \mathcal{I}_\eta=4\left[\langle \partial_\eta \phi_r |\partial_\eta \phi_r \rangle+(\langle \partial_\eta \phi_r |\phi_r\rangle)^2 \right], 
    \end{aligned}
\end{equation}
where $|\partial_\eta \phi_r \rangle \equiv \partial |\phi_r\rangle / \partial \eta$ denotes the derivative of the photonic mode’s ground state with respect to the parameter $\eta$. 
Therefore, the QFI for the photonic mode is given by
\begin{equation}
    \begin{aligned}        
        \mathcal{I}_\eta={}&\frac{\eta^2}{2(1-\eta^2)^2}. \label{eq:QFI}
    \end{aligned}
\end{equation}
As shown in Fig.~\ref{fig:inverted_variance_eta}, $\mathcal{I}_\eta$ exhibits a divergent behavior as $\eta \to 1$, reflecting a critical enhancement of the quantum probe’s sensitivity near the phase transition.  
This divergence implies that, in principle, the estimation precision can be dramatically improved by operating close to the critical point.
The fundamental limit on the achievable variance in any unbiased estimator of $\eta$ is set by the quantum Cram{\'e}r-Rao bound
\begin{equation}
    \begin{aligned}        
        \delta^2 \eta \ge (\nu \mathcal{I}_\eta)^{-1},
    \end{aligned}
\end{equation}
where $\nu$ denotes the number of independent measurement repetitions and the bound assumes access to an optimal measurement strategy.  
While the QFI quantifies the ultimate theoretical sensitivity, its practical relevance hinges on the existence of a physically realizable measurement scheme that can saturate or approach this bound.  
In what follows, we present experimentally feasible protocols capable of attaining an estimation precision that scales identically to the quantum Cram{\'e}r-Rao bound, thereby harnessing the full metrological advantage offered by critical quantum dynamics.

\section{Measurement protocol}

To ensure the system remains in its instantaneous eigenstate throughout the dynamics, we employ an adiabatic evolution strategy. Starting from the prepared dark eigenstate, the system parameter ($\eta$) is varied sufficiently slowly such that the state adiabatically tracks the ground state of the time-dependent Hamiltonian without populating excited states. 
After evolving for a duration $t$ under the system Hamiltonian given in Eq.~(\ref{eq:H}), we perform a projective measurement of the average photon number $\langle N \rangle$, where the photon number operator is defined as $N = a^\dagger a$. 
Its expectation value and fluctuations are found to be
\begin{subequations}
    \begin{align}
        \langle N\rangle_t={}&\frac{2-\eta_t^2}{4\sqrt{1-\eta_t^2}}-\frac{1}{2}, \label{eq:average photon number}\\
        (\Delta N)^2_t={}&\frac{(1-\eta_t^2)^2+1}{8(1-\eta_t^2)}-\frac{1}{4},
    \end{align}
\end{subequations}
where all quantities with subscript $t$ are evaluated at time $t$.
To evaluate the metrological utility of this observable, we compute its susceptibility with respect to the control parameter:
\begin{equation}
    \begin{aligned}
        \chi_\eta(t)=\frac{\partial \langle N\rangle_t}{\partial \eta_t}=\frac{\eta_t ^3}{4 \left(1-\eta_t ^2\right)^{3/2}},\\
    \end{aligned}
\end{equation}
which diverges algebraically as $\eta_t \to 1$, a hallmark of criticality that underpins the enhanced sensitivity of quantum sensors operating close to the critical point.
The ultimate precision achievable in estimating $\eta_t$ is bounded by the quantum Cram{\'e}r-Rao bound, and for a given measurement strategy, the relevant figure of merit is the signal-to-noise ratio squared, often referred to as the inverted variance:
\begin{equation}
    \begin{aligned}
        \mathcal{F}_\eta(t)=\frac{\chi^2_\eta(t)}{(\Delta N)^2_t}=\frac{\eta_t ^2}{2\left(1-\eta_t ^2\right)^2}.\\
    \end{aligned}
\end{equation}
Remarkably, this expression coincides exactly with the quantum Fisher information associated with the dark eigenstate, indicating that photon-number detection is one of the optimal measurements for this encoding scheme and saturates the fundamental quantum limit.
In the asymptotic regime close to criticality, where $1 - \eta_t^2 \ll 1$, one may approximate $\langle N \rangle_t \sim (4\sqrt{1 - \eta_t^2})^{-1}$ and combine it with the time dependence in Eq.~(\ref{eq:varepsilon_t}) to obtain the scaling law (see Section S3 of Supporting Information for details)
\begin{equation}
    \begin{aligned}
        \mathcal{F}_\eta(t) \propto \langle N \rangle_t t^2.\\
    \end{aligned}
\end{equation}
This dual dependence reveals that the estimation precision benefits simultaneously from both the time to encode the signal and the photon number of the field mode.
Consequently, the protocol achieves the Heisenberg-limited scaling, where the uncertainty in estimating $\eta$ scales as $\delta \eta \sim 1/(\sqrt{\langle N \rangle_t}\, t)$, surpassing the standard quantum limit and approaching the ultimate bound allowed by quantum mechanics.

Beyond photon-number detection, alternative measurement strategies based on field quadratures offer complementary routes to high-precision sensing.
In particular, one may measure the second-moment statistics of the canonical quadrature operators $X = (a^\dagger + a)/2$ and $P = i(a^\dagger - a)/2$, which are directly accessible via homodyne detection in microwave or optical quantum systems.
For the position-like quadrature $X$, the mean square and the variance of $X^2$ are given by
\begin{equation}
    \begin{aligned}
        \langle X^2\rangle_t={}&\frac{1}{4}(1-\eta_t^2)^{-1/2},\\
        (\Delta X^2)_t^2={}&\frac{1}{8}(1-\eta_t^2)^{-1},\\        
    \end{aligned}
\end{equation}
while for the momentum-like quadrature $P$, we find
\begin{equation}
    \begin{aligned}
        \langle P^2\rangle_t={}&\frac{1}{4}(1-\eta_t^2)^{1/2},\\
        (\Delta P^2)_t^2={}&\frac{1}{8}(1-\eta_t^2).\\
    \end{aligned}
\end{equation}
These results reflect the squeezing-induced asymmetry between conjugate quadratures: as the system approaches the critical point, quantum fluctuations in $X$ are dramatically amplified, whereas those in $P$ are suppressed.
Despite this asymmetry, the metrological gain derived from either quadrature yields the same sensitivity. 
Specifically, using the response of $\langle X^2 \rangle_t$ (or $\langle P^2 \rangle_t$) to variations in $\eta_t$, the corresponding inverted variance is
\begin{equation}
    \begin{aligned}
        \mathcal{F}_\eta(t)={}&\frac{(\partial_\eta \langle X^2\rangle_t)^2}{{\rm Var}[X^2]}=\frac{\eta_t ^2}{2\left(1-\eta_t ^2\right)^2},
    \end{aligned}
\end{equation}
and analogously for $P$,
\begin{equation}
    \begin{aligned}
        \mathcal{F}_\eta(t)={}&\frac{(\partial_\eta \langle P^2\rangle_t)^2}{{\rm Var}[P^2]}=\frac{\eta_t ^2}{2\left(1-\eta_t ^2\right)^2}.
    \end{aligned}
\end{equation}
Thus, both quadrature-based protocols also attain the Heisenberg limit in the critical regime, providing experimentally flexible alternatives that do not require direct photon counting.
This diversity across multiple observables underscores the universality of critical enhancement in quantum metrology and highlights the versatility of the driven JCM for high-precision sensing applications.

While both the present protocol and our prior work \cite{Lue2026Critical} exploit criticality for quantum-enhanced sensing, 
they differ fundamentally in the choice of observable and its robustness to noise. 
In Ref. \cite{Lue2026Critical}, the signal was encoded in the qubit excitation number, which approaches $1/2$ for all eigenstates as $\eta_t \to 1$, 
making it insensitive to leakage and thus inherently robust. 
By contrast, the current scheme utilizes the photon number ($N$) and the square of the quadrature operators ($X^2$ and $P^2$), whose inverted variances diverge near criticality, enabling Heisenberg-limited scaling in both time and photon number. 
However, this sensitivity comes at the cost of fragility to leakage into bright states $|\psi_{n,\pm}\rangle$, 
whose expectation values for all three observables deviate significantly from those of the dark state, thereby degrading the signal-to-noise ratio. 
Hence, unlike the qubit-based estimator, the field-based estimator is not intrinsically noise-resilient, as such an approach prioritizes maximal sensitivity under controlled conditions.

\section{Adiabatic Evolution}

In our quantum sensing protocol, the signal of interest is encoded into the time-evolving dark eigenstate $\left\vert \psi _{0}\right\rangle$, i.e., the ground state, which is separated from the nearest excited eigenstates by a gap $\Delta E = \Omega (1 - \eta^{2})^{3/4}$.
The integrity of the adiabatic evolution—i.e., the ability of the system to remain in the instantaneous dark state throughout the protocol—is determined by the interference between this energy gap and the rate at which $\eta$ is varied in time. 
Specifically, slower ramping ensures better adiabatic following, while rapid changes may induce non-adiabatic transitions that degrade sensing sensitivity.
The temporal trajectory of the control parameter is explicitly prescribed as (see Section S2 of Supporting Information for details)
\begin{equation}
    \begin{aligned}        
        \eta_t = \sqrt{1 - [(k t)^{4/3} + 1]^{-1}}, \label{eq:varepsilon_t}
    \end{aligned}
\end{equation}
where $k > 0$ is a tunable coefficient that governs the effective ramping velocity. 
The exponent $4/3$ is chosen to mitigate critical slowing down near the quantum phase transition: it ensures that the  $\dot{\eta}_t$ scales in a way that largely compensates for the vanishing energy gap, thereby suppressing non-adiabatic excitations and preserving high adiabatic fidelity throughout the evolution.

\begin{figure}[t]
    \centering
    \includegraphics[width=\linewidth]{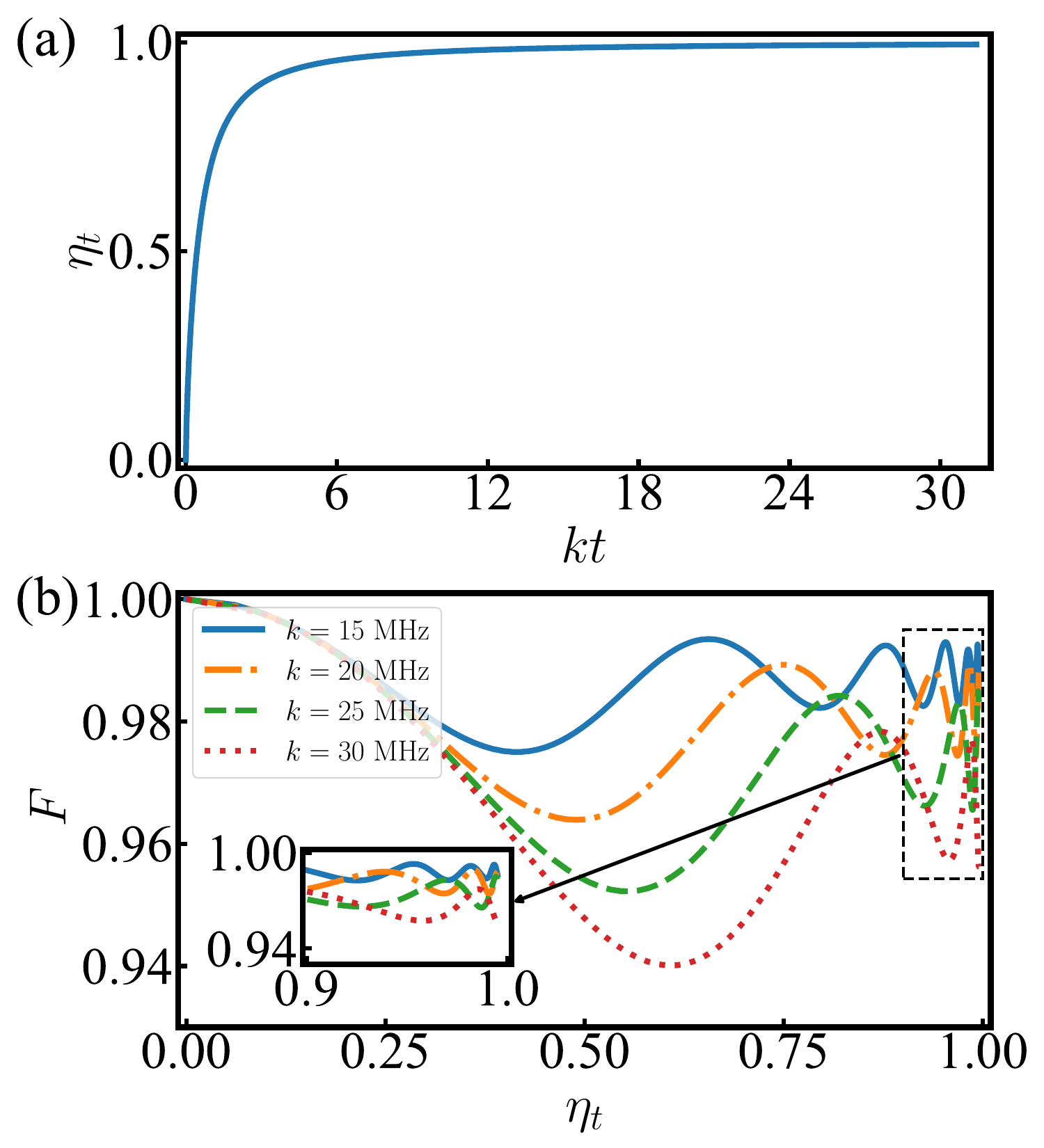}
    \caption{
        (a) Dependence of $\eta$ on the dimensionless time variable $kt$.  
        (b) Fidelity $F$ of the qubit-photon state with respect to the ideal dark state as a function of $\eta_t$. 
        The inset provides a magnified view of the regime $0.90 < \eta_t < 0.995$.
        The simulation is performed under unitary dynamics with a qubit-resonator coupling strength of $\Omega/2\pi = 20.9 \ \text{MHz}$. At $\eta_t = 0.995$, the fidelities are 0.986, 0.988, 0.983, and 0.956 for ramping rates \(k = 15\), \(20\), \(25\), and \(30~\text{MHz}\), respectively.
    }
    \label{fig:fig_fidelity_eta_nodecay}
\end{figure}

This functional form is carefully chosen to interpolate smoothly between the non-driven regime ($\eta_t \approx 0$ at early times) and the critical region ($\eta_t \to 1$ as $t \to \infty$), thereby enabling controlled access to the vicinity of the quantum critical point without abrupt parameter jumps.
The rescaled field amplitude $\eta_t$, plotted as a function of the dimensionless time variable $kt$, is shown in Fig.~\ref{fig:fig_fidelity_eta_nodecay}(a). 
Larger value of $k$ can accelerate the approach to the critical regime, reducing the time required to reach a fixed target value such as $\eta = 0.995$; however, this comes at the cost of increased non-adiabatic excitations, highlighting the fundamental trade-off between sensing speed and adiabatic fidelity in critical quantum metrology.

The theoretical analysis discussed so far assumes perfect isolation of the system from its environment. To evaluate the performance of our protocol under idealized yet physically insightful conditions, we consider unitary dynamics governed solely by the time-dependent Hamiltonian $H_I(t)$ given in Eq.~(\ref{eq:H}). Under these conditions, the state evolution is described by the Schr\"odinger equation,
\begin{equation}
    \begin{aligned}
        i \frac{d}{dt} |\Psi(t)\rangle = H_I(t) |\Psi(t)\rangle,
    \end{aligned}
\end{equation} 
with the system initialized in the instantaneous dark eigenstate $|\psi_0(0)\rangle$ at $t = 0$.

We numerically simulate the unitary quantum dynamics governed by the Schrödinger equation as $\eta_t$ approaches the critical point according to Eq.~(\ref{eq:varepsilon_t}). The fidelity between the evolved state $|\Psi(t)\rangle$ and the instantaneous ideal dark state $|\psi_0(t)\rangle$ is then computed as $F(t) = |\langle \psi_0(t) | \Psi(t) \rangle|^2$. 
The fidelity \(F\) of the evolved qubit-photon state with respect to the ideal dark state is shown as a function of the control parameter \(\eta_t\) in Fig.~\ref{fig:fig_fidelity_eta_nodecay}(b). At \(\eta_t = 0.995\), the fidelities are \(0.986\), \(0.988\), \(0.983\), and \(0.956\) for ramping rates \(k = 15\), \(20\), \(25\), and \(30~\text{MHz}\), respectively. The drop in fidelity as \(\eta_t\) approaches 1 is due to non-adiabatic excitations that grow stronger when the energy gap becomes small. However, this effect can be reduced by choosing an appropriate ramping rate. These results show that the target dark state can be reliably reached under realistically controllable driving conditions while  approaching the critical point.
A detailed discussion of the impact of decoherence on qubit-photon state fidelity is provided 
in Section S4 of Supporting Information.

\section{CONCLUSIONS}
In conclusion, we have proposed a quantum sensing protocol by making use of
the critical behaviors of a photonic mode, which interacts with a qubit by
photonic swapping. The external signal field, whose strength is to be
estimated, is coupled to the photonic mode. Following the dark eigenstate,
the photonic field remains in a squeezed vacuum state, with the squeezing
parameter displaying a diverging behavior at the critical point and thus usable to amplify the signal. The measurement precision of the critical
sensor exhibits a Heisenberg scaling with respect to the evolution time and
the photon number under the ideal conditions. We further investigate the
performance of the sensor including decoherence effects. The
scheme serves as a complement to the previous ones that are also based on
the divergent behaviors of squeezed vacuum states in qubit-photon systems,
but focus on measurement of the inherent properties of the quantum system
\cite{Garbe2020Critical,Chu2021Dynamic,Zhu2023Criticality,Lue2022Critical}.

\section{ACKNOWLEDGEMENTS}
This work was supported by the National Natural Science Foundation of China (Grants No. 12274080, No. 12474356, No. 12475015, and No. 12505016), Quantum Science and Technology-National Science and Technology Major Project (Grant No. 2021ZD0300200), and the Natural Science Foundation of Fujian Province (Grant No. 2025J01465). 

\section{Conflict of Interest}
The authors declare that they have no conflict of interest.

\section{Supporting Information}
The supporting information is available online at \url{http://phys.scichina.com}
and \url{https://link.springer.com}. The supporting materials are published as
submitted, without typesetting or editing. The responsibility for scientific
accuracy and content remains entirely with the authors.

\bibliography{ref} 
\end{document}


\title{Supplementary Materials for \\``Quantum-enhanced estimation of signal field amplitudes with critical squeezed states of photonic modes"}

\author{Ken Chen}
\author{Jia-Hao L\"{u}}
\author{Wen Ning} \email{ningw@fzu.edu.cn}
\address{Fujian Key Laboratory of Quantum Information and Quantum\\
Optics, College of Physics and Information Engineering,\\
Fuzhou University, Fuzhou, Fujian 350108, China}

\author{Zhen-Biao Yang} \email{zbyang@fzu.edu.cn}
\author{Shi-Biao Zheng} \email{t96034@fzu.edu.cn}
\address{Fujian Key Laboratory of Quantum Information and Quantum\\
Optics, College of Physics and Information Engineering,\\
Fuzhou University, Fuzhou, Fujian 350108, China}
\address{Hefei National Laboratory, Hefei 230088, China}

\maketitle

\tableofcontents

\section{DERIVATION OF THE QFI}
We here provide the detailed derivation of the QFI for the photonic mode's ground state $|\phi_r\rangle = S(r)|0\rangle$, where $S(r) = \exp[r(a^2 - a^{\dagger 2})/2]$ is the squeezing operator and the squeezing parameter is $r = \frac{1}{4}\ln(1-\eta^2)$.
The QFI for a pure state is given by $\mathcal{I}_\eta=4\left[\langle \partial_\eta \phi_r |\partial_\eta \phi_r \rangle+(\langle \partial_\eta \phi_r |\phi_r\rangle)^2 \right]$. We first compute the derivative of the state with respect to $\eta$:
\begin{equation}
    \begin{aligned}
|\partial_\eta \phi_r \rangle = \frac{\partial S(r)}{\partial \eta}|0\rangle = \frac{\eta}{4(1-\eta^2)} (a^{\dagger 2} -a^2 ) S(r) |0\rangle.
    \end{aligned}
\end{equation}
We then evaluate the two terms in the QFI formula. The first term is
\begin{equation}
    \begin{aligned}
\langle \partial_\eta \phi_r | \phi_r \rangle = \frac{\eta}{4(1-\eta^2)} \langle 0 | (a^2 - a^{\dagger 2}) | 0 \rangle = 0.
    \end{aligned}
\end{equation}
For the second term, we calculate
\begin{equation}
    \begin{aligned}
        \langle \partial_\eta \phi_r | \partial_\eta \phi_r \rangle ={}& \frac{\eta^2}{16(1-\eta^2)^2} \langle 0 | (a^2 - a^{\dagger 2})^2 | 0 \rangle\\
        ={}&\frac{\eta^2}{8(1-\eta^2)^2}.
    \end{aligned}
\end{equation}
Therefore, the QFI is given by
\begin{equation}
    \begin{aligned}
\mathcal{I}_\eta ={}& 4\left[\langle \partial_\eta \phi_r |\partial_\eta \phi_r \rangle+(\langle \partial_\eta \phi_r |\phi_r\rangle)^2 \right]\\
={}& \frac{\eta^2}{2(1-\eta^2)^2},
    \end{aligned}
\end{equation}
which is Eq. (7) in the main text.

\section{ADIABATIC EVOLUTION}\label{appendix:eta_time_function}

We consider a quantum sensing protocol in which the system is initially prepared in the instantaneous dark eigenstate $\left\vert \psi_0 \right\rangle$.\
As the dimensionless control parameter $\eta$ is varied slowly in time, non-adiabatic transitions to excited eigenstates $\left\vert \psi_{n,\pm} \right\rangle$ may occur.\
Within the framework of first-order time-dependent perturbation theory, the transition probability from the ground state to an excited state $\left\vert \psi_{n,\pm} \right\rangle$ scales as
\begin{equation}
    \begin{aligned}
        P_{n,\pm }\sim \left\vert \frac{\left\langle \psi _{n}^{\pm }\right\vert 
        \dot{H}_{I}\left\vert \psi _{0}\right\rangle }{(E_{n}^{\pm })^{2}}
        \right\vert ^{2},
    \end{aligned}
\end{equation}
where $\dot{H}_{I} = dH_{I}/dt$ denotes the time derivative of the system Hamiltonian, and $E_{n,\pm}$ is the energy gap between the dark state and the target excited state.

Given the explicit form of the Hamiltonian, its time dependence enters solely through $\eta(t)$, yielding
\begin{equation}
    \begin{aligned}
        \dot{H}_{I}=\dot{\eta }\Omega (a^{\dagger}+a)/2.
    \end{aligned}
\end{equation}
with $\Omega$ representing the coupling strength associated with the drive.
Using the identity for the squeezing operator $S(r )=\exp [r(a^{2}- a^{\dagger 2})/2]$, 
\begin{equation}
    \begin{aligned}
        S^\dagger(r)(a^{\dagger }+a)S(r)=e^{-r}(a^{\dagger }+a),
    \end{aligned}
\end{equation}
and recalling that the eigenstates $\left\vert \psi_{n,\pm} \right\rangle$ involve both displacement $D(\alpha_{n,\pm})$ and squeezing $S(r)$ operations, we evaluate the matrix element to obtain
\begin{equation}
    \begin{aligned}
        P_{n,\pm} \sim 
        \left|\frac{\dot{\eta} e^{-r} (-\eta\langle n-1| \pm \langle n|)   (a^{\dagger }+\alpha_{n,\pm}^*) D^\dagger(\alpha_{n,\pm})|0 \rangle}{2\sqrt{2} n\Omega (1-\eta^2)^{3/2}}   \right|^2. \label{eq:Pn}
    \end{aligned}
\end{equation}
Substituting the explicit expressions for the squeezing parameter $r = \frac{1}{4} \ln(1 - \eta^2)$ and the displacement amplitude $\alpha_{n,\pm} = \mp \sqrt{n}\, \eta$, and evaluating the resulting Fock-state overlaps, the transition probability simplifies to
\begin{equation}
    \begin{aligned}
        P_{n,\pm} \sim \left|\frac{\dot{\eta}  \eta e^{-n\eta^2/2} (\pm \sqrt{n}\eta)^{n-2} }{2\sqrt{2} n\Omega (1-\eta^2)^{7/4} \sqrt{(n-1)!} } \right|^2\\          
    \end{aligned}
\end{equation}

For the lowest excited eigenstate ($n=1$), this expression reduces significantly as follows:
\begin{equation}
    \begin{aligned}
        P_{1,\pm } &\sim \left|\frac{\dot{\eta}  e^{-\eta^2/2} }{2\sqrt{2} \Omega (1-\eta^2)^{7/4} }\right|^2.
    \end{aligned}
\end{equation}
To analyze the dynamics near the critical point, we adopt the following time-dependent protocol for the control parameter:
\begin{equation}
    \begin{aligned}
        \eta =\sqrt{1-[(kt)^{\xi}+1]^{-1}}. 
    \end{aligned}
\end{equation}
where $k > 0$ sets the effective ramping rate and $\xi > 0$ is an exponent that governs the approach to criticality.
In the asymptotic regime $k t \gg 1$ (i.e., close to the critical point $\eta \to 1$), we may approximate
\begin{equation}
    \begin{aligned}
        \eta(t) \simeq 1 - \frac{1}{2(k t)^{\xi}} \simeq \exp \left[ -\frac{1}{2(k t)^{\xi}} \right],
    \end{aligned}
\end{equation}
which leads to the time derivative
\begin{equation}
    \begin{aligned}
        \dot{\eta } \simeq \frac{\xi }{2k^{\xi }t^{\xi +1}}
        e^{-1/[2(kt)^{\xi}]} 
        \simeq \frac{\eta \xi k}{2}(1-\eta ^{2})^{(\xi +1)/\xi }.
    \end{aligned}
\end{equation}
Inserting this expression for $\dot{\eta}$ into the formula for $P_{1,\pm}$ yields 
\begin{equation}
    \begin{aligned}
        P_{1,\pm }\sim \left|\frac{\eta \xi k   e^{-\eta^2/2}  (1-\eta^2)^{(\xi+1)/\xi -7/4}}{4\sqrt{2} \Omega  }\right|^2. 
    \end{aligned}
\end{equation}

Choosing the exponent $\xi = 4/3$, which is motivated by critical slowing down and optimal adiabatic passage in similar driven-dissipative models, the exponent of $(1 - \eta^2)$ vanishes.
Consequently, the transition probability becomes approximately constant near criticality:
\begin{equation}
    \begin{aligned}
        P_{1,\pm }\sim \left(\frac{\eta  k   e^{-\eta^2/2}}{3\sqrt{2} \Omega}\right)^2 \simeq \left(\frac{  k   e^{-1/2}}{3\sqrt{2} \Omega}\right)^2,
    \end{aligned}
\end{equation}
where we used $\eta \to 1$ in the final approximation.
Thus, provided the ramping rate satisfies $k < \Omega$, we ensure $P_{1,\pm} \ll 1$, validating the adiabatic approximation for the lowest excitation channel.

For higher excited states ($n > 1$), using the same choice $\xi = 4/3$ and evaluating the asymptotic form near $\eta \to 1$, we find
\begin{equation}
    \begin{aligned}
        P_{n,\pm} &\sim \left( \frac{ k\, e^{-n/2} (\pm \sqrt{n})^{\,n-2} }{ 3\sqrt{2}\, n \Omega \sqrt{(n-1)!} } \right)^2 \\
        &= \left( \frac{ e^{-n/2} n^{(n-3)/2} }{ \sqrt{n!} } \cdot \frac{k}{3\sqrt{2}\, \Omega} \right)^2.
    \end{aligned}
\end{equation}
Applying Stirling's approximation $n! \sim \sqrt{2\pi n}\, (n/e)^n$, one can verify that the prefactor
\[
\frac{ e^{-n/2} n^{(n-3)/2} }{ \sqrt{n!} } \sim \mathcal{O}(n^{-1})
\]
decays rapidly with $n$. Hence, for all $n \geq 1$,
\begin{equation}
    \begin{aligned}
        P_{n,\pm} < \left( \frac{k}{3\sqrt{2}\, \Omega} \right)^2.
    \end{aligned}
\end{equation}
Therefore, under the condition $k < \Omega$, all non-adiabatic transition probabilities remain negligible throughout the evolution.
This confirms that the system remains confined to the dark-state manifold with high fidelity, justifying the adiabatic encoding assumption employed in the main text.

\section{SCALING TOWARD THE HEISENBERG LIMIT}\label{eq:Heisenberg_limit}
To elucidate the metrological scaling of our protocol near the quantum critical point, we introduce the small parameter $\epsilon = 1 - \eta^2$, which quantifies the distance from criticality ($\eta \to 1$ corresponds to $\epsilon \to 0$).
Using the prescribed time dependence of the control parameter,
\begin{equation}
    \begin{aligned}
        \eta(t) = \sqrt{1 - \big[(k t)^{4/3} + 1\big]^{-1}},
    \end{aligned}
\end{equation}
we obtain the exact relation
\begin{equation}
    \begin{aligned}
        \epsilon(t) = \big[(k t)^{4/3} + 1\big]^{-1}.
    \end{aligned}
\end{equation}
In the asymptotic regime close to the critical point—i.e., for sufficiently long evolution times such that $k t \gg 1$—this simplifies to the power-law decay
\begin{equation}
    \begin{aligned}
        \epsilon \sim (k t)^{-4/3}.
    \end{aligned}
\end{equation}

The inverted variance with respect to the parameter $\epsilon$, denoted $\mathcal{F}_\epsilon(t)$, governs the ultimate precision limit for estimating $\epsilon$.
Starting from the expression derived in the main text,
\begin{equation}
    \begin{aligned}
        \mathcal{F}_\eta(t) = \frac{\eta^2}{2(1 - \eta^2)^2} = \frac{1 - \epsilon}{2 \epsilon^2},
    \end{aligned}
\end{equation}
we find that, in the critical vicinity where $\epsilon \ll 1$,
\begin{equation}
    \begin{aligned}
        \mathcal{F}_\epsilon(t) = \frac{1 - \epsilon}{2 \epsilon^2} \sim \frac{1}{2 \epsilon^2} \sim \frac{1}{2} (k t)^{8/3}.
    \end{aligned}
\end{equation}
This super-linear growth with time already indicates enhanced sensitivity beyond the standard quantum limit.

To connect this scaling to physical resources, we re-express the average photon number $\langle N \rangle_t$, originally given in Eq.~(9a) of the main text, in terms of $\epsilon$:
\begin{equation}
    \begin{aligned}
        \langle N\rangle_t &= \frac{1}{4\sqrt{\epsilon}} + \frac{\sqrt{\epsilon}}{4} - \frac{1}{2}.
    \end{aligned}
\end{equation}
Near the critical point ($\epsilon \to 0$), the dominant contribution arises from the first term, yielding the asymptotic behavior
\begin{equation}
    \begin{aligned}
        \langle N\rangle_t \sim \frac{1}{4\sqrt{\epsilon}} \simeq \frac{1}{4} (k t)^{2/3}.
    \end{aligned}
\end{equation}
Thus, the average photon number—the primary resource cost in microwave or optical quantum systems—grows sublinearly with time.

Crucially, combining the above scalings reveals a fundamental metrological relation.
Substituting $\epsilon \sim (k t)^{-4/3}$ into $\mathcal{F}_\eta(t) \sim 1/(2\epsilon^2)$ and using $\langle N \rangle_t \sim 1/(4\sqrt{\epsilon})$, we eliminate $\epsilon$ to obtain
\begin{equation}
    \begin{aligned}
        \mathcal{F}_\eta(t) \propto \langle N \rangle_t \, t^2.
    \end{aligned}
\end{equation}
This proportionality signifies that the inverted variance scales quadratically with the time and linearly with the photon number of the field mode.
Consequently, the estimation uncertainty for the parameter $\eta$ obeys
\begin{equation}
    \begin{aligned}
        \delta \eta \geq \frac{1}{\sqrt{\mathcal{F}_\eta(t)}} \propto \frac{1}{\sqrt{\langle N \rangle_t}\, t},
    \end{aligned}
\end{equation}
which saturates the Heisenberg limit—a hallmark of optimal quantum-enhanced metrology.
This result underscores that criticality not only amplifies susceptibility but also enables resource-efficient sensing, where temporal and photonic resources jointly achieve the ultimate quantum bound.

\begin{figure}[h]
    \centering
    \includegraphics[width=0.77\linewidth]{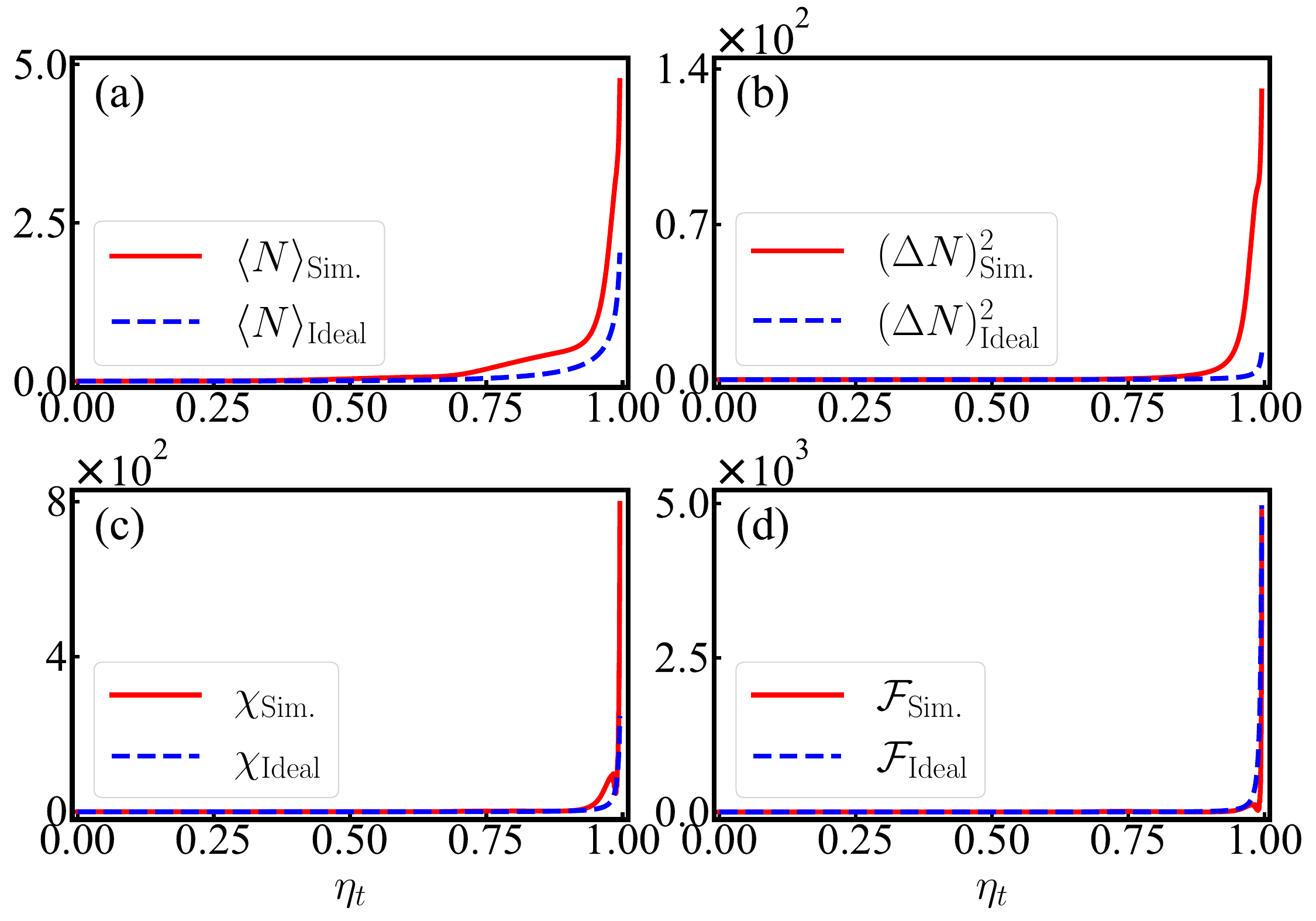}
    \caption{
        The influences of decoherence on the sensing protocol under the experimentally accessible conditions. 
        (a) The average photon number, (b) the variance of the photon number, (c) the susceptibility, and (d) the inverted variance, as a function of $\eta_t \in [0, 0.995]$, respectively. In all panels, the dashed (solid) line denotes the ideal (simulated) case with experimentally accessible parameters \cite{Lue2026Critical}: $\Omega/2\pi = 20.9 \ \text{MHz}$, $T_{1,q} = 20$ $\upmu$s, and $T_{1,r} = 12$ $\upmu$s; $T_{q,\varphi}$ and $T_{r,\varphi}$ are ignored by considering that the continuous driving dynamics greatly prolonging such dephasing times \cite{Lue2026Critical,Guo2018Dephasing}. The ramping rate $k = 34.5$ MHz is set.   
    }
    \label{fig:fig_F_N_k34_T1q20_T2q0_T1r12}
\end{figure}

\section{THE INFLUENCES OF DECOHERENCE}\label{eq:decoherence}
The results presented in the main text assume an isolated system. To evaluate the feasibility of our protocol under realistic, noisy conditions, we analyze its performance using a master equation approach that incorporates the dominant decoherence channels. The dynamics of the system is governed by 
\begin{equation}
    \begin{aligned}
        \dot{\rho}={}&-i[H_I,\rho]+\frac{1}{T_{1,q}} \mathcal{L}[\sigma_-]\rho + \frac{1}{T_{\varphi,q}} \mathcal{L}[\sigma_z]\rho\\
        {}& + \frac{1}{T_{1,r}} \mathcal{L}[a]\rho + \frac{1}{T_{\varphi,r}} \mathcal{L}[a^\dagger a]\rho,
    \label{master}
    \end{aligned}
\end{equation}
where the Lindblad super-operator is defined as $\mathcal{L}[A]\rho=A\rho A^\dagger-A^\dagger A\rho/2-\rho A^\dagger A/2$ for any dissipator operator $A$ (e.g., $\sigma_-,\ \sigma_z,\ a,\ a^\dagger a$). Consistent with typical operating conditions for platforms like superconducting circuits at millikelvin temperatures \cite{Lue2026Critical,Xu2023Digital}, thermal excitations of the qubit and resonator are negligible. Here, $\sigma_z$ is the qubit inversion operator defined as $\sigma_z=\sigma_+\sigma_- - \sigma_-\sigma_+$, with $\sigma_+$ ($\sigma_-$) being the qubit raising (lowering) operator, $T_{1,q}$ ($T_{\varphi,q}$) and $T_{1,r}$ ($T_{\varphi,r}$) are the qubit energy relaxation (dephasing) and resonator photon energy relaxation (dephasing) times, respectively.

\begin{figure}[htbp]
    \centering
    \includegraphics[width=0.8\linewidth]{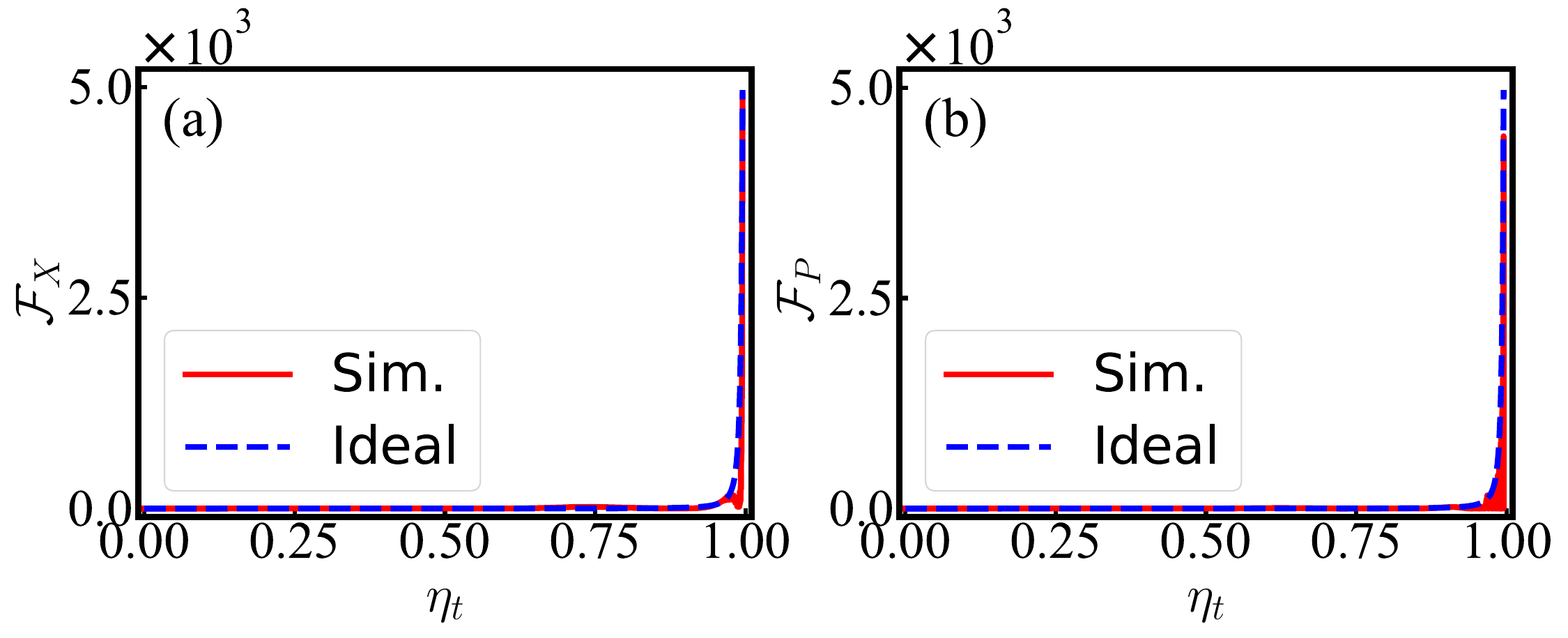}
    \caption{
         The inverted variance of the quadrature operators $X$ (a) and $P$ (b) with respect to $\eta_t \in [0, 0.995]$, respectively.
         Simulations are performed with experimentally accessible parameters \cite{Lue2026Critical}: $\Omega/2\pi = 20.9 \ \text{MHz}$, $T_{1,q} = 20$ $\upmu$s, $T_{1,r} = 12$ $\upmu$s. The ramping rates are $k=34.2$ MHz and $k=23$ MHz, respectively.
    }
    \label{fig:fig_F_Xk34.2_Pk23_T1q20_T2q0_T1r12_eta_0.995}
\end{figure}

Following the experimental dynamics where continuous driving plays a pivotal role \cite{Guo2018Dephasing,Lue2026Critical}, dephasing of both the  
qubit and the resonator is ignored. 
In Fig.~\ref{fig:fig_F_N_k34_T1q20_T2q0_T1r12}, we plot the average photon number (a), the variance of the photon number (b), the susceptibility (c), and the inverted variance (d) with respect to $\eta_t$, respectively, under the influences of decoherence. The numerical calculation uses the experimentally accessible parameters \cite{Lue2026Critical}: $\Omega/2\pi = 20.9$ MHz, $T_{1,q} = 20 \ \upmu$s, $T_{1,r} = 12 \ \upmu$s. 
As shown in Fig.~\ref{fig:fig_F_N_k34_T1q20_T2q0_T1r12}(d), 
the protocol remains effective under realistic decoherence, and the inverted variance approaches the Heisenberg scaling, indicating quantum-enhanced sensing. 
Moreover, Fig.~\ref{fig:fig_F_Xk34.2_Pk23_T1q20_T2q0_T1r12_eta_0.995} shows that the numerical results for the inverted variances of 
$X$ and $P$ remain in close agreement with the ideal case, even in the presence of decoherence.

\bibliography{ref_sub}